\documentstyle[aps,12pt]{revtex}
\def \to {\rightarrow}
\def \beq {\begin{equation}}
\def \eeq {\end{equation}}
\def \ba {\begin{eqnarray}}
\def \ea {\end{eqnarray}}
\def \jpsi {J/\psi}
\def \ee {e^+e^-}
\begin{document}
\baselineskip 20pt
\renewcommand{\thesection}{\Roman{section}}
~
\vbox{\halign{&#\hfil\cr
                & BUTP-96-30   \cr
                & October 1996  \cr}
                }
\vskip 15mm
\begin{center}
{\Large \bf Prompt $J/\psi$ production at $e^+e^-$ colliders} 
\end{center}
\vskip 10mm
\centerline{Feng Yuan}
\vskip 2mm
\centerline{\small\it Department of Physics, Peking University, 
            Beijing 100871, People's Republic of China}
\vskip 4mm
\centerline{Cong-Feng Qiao}
\vskip 2mm
\centerline{\small\it China Center of Advanced Science and Technology
(World Laboratory), Beijing 100080, People's Republic of China}
\vskip 4mm
\centerline{Kuang-Ta Chao}
\vskip 2mm
\centerline{\small\it China Center of Advanced Science and Technology
(World Laboratory), Beijing 100080, People's Republic of China}
\vskip 1mm
\centerline{\small\it Department of Physics, Peking University, 
            Beijing 100871, People's Republic of China}
\vskip 15mm

\begin{center}

{\bf\large Abstract}

\begin{minipage}{140mm}
\vskip 5mm
\par
In this paper, we  discuss the prompt $\jpsi$ production at $\ee$ colliders
via color-singlet and color-octet production mechanisms. The color-singlet
production processes include 1) $e^+e^-\to J/\psi gg$; 2) $e^+e^-\to J/\psi
c\bar{c}$; 3) $e^+e^-\to q \bar q ggJ/\psi$ and $e^+e^-\to q \bar q g\chi_c$
followed by $\chi_c \to J/\psi \gamma$. The color-octet production processes
include 1) $e^+e^-\to J/\psi g$; 2) $\ee\to \jpsi q\bar q$. Of all these
production channels, we find that the color-octet contributions dominate over
the color-singlet contributions at any energy scales. At low energies
($\sqrt{s}< 20 GeV$), the dominant channel is $e^+e^-\to J/\psi g$ whereas
at high energies $\ee\to \jpsi q\bar q$ will take the leading part. We also
find that the energy spectrum for the color-octet $\jpsi$ production in
process $\ee\to \jpsi q\bar q$ is very soft, and the mean energy of the
produced $\jpsi$ is only about $10 GeV\sim 20 GeV$ even at very high energies
(e.g. at $1000 GeV$). The extraction of color-octet matrix elements from
$\jpsi$ production in $\ee$ collisions is also discussed.

\vskip 5mm
\noindent
PACS number(s): 13.60.Le, 13.20.Gd
\end{minipage}
\end{center}
\vfill\eject\pagestyle{plain}\setcounter{page}{1}

\section{Introduction}
~

Since the first charmonium state $\jpsi$ was discovered in 1974, the studies of 
heavy quarkonium states (include charmonium and bottomonium) have played an important 
role in elementary particle physics. Charmonium and bottomonium are the
simplest quark-antiquark composite particles,
which can be described by the gauge theory Quantum Chromodynamics (QCD). The 
investigation of their properties, such as the production and decays 
may help us to reveal the properties of QCD in both perturbative and 
nonperturbative sectors. In particular, the $\jpsi$ production 
is of special significance because it has extremely clean signature through its
leptonic decay modes. Therefore if the heavy quarkonium production mechanism is clarified, $\jpsi$ 
triggers can be used as a powerful tool in studying other interesting physics.

The studies of heavy quarkonium production in recent years are mainly stimulated
by large discrepancies between the color-singlet model (CSM) predictions and the recent 
experimental data of {\bf CDF} at {\bf Fermilab Tevatron}\cite{cdf}. 
There is orders of magnitude disagreement between them.
In the color-singlet model description for the quarkonium states,
the quark-antiquark pairs are created in colorless configurations.
In the past few years, a rigorous framework for treating quarkonium production 
and decays has been advocated by Bodwin, Braaten and Lepage in the context of 
nonrelativistic quantum chromodynamics (NRQCD) \cite{bbl}. 
In this approach, the production process is factorized into short and long 
distance parts, while the latter is associated with the nonperturbative
matrix elements of four-fermion operators.
Another outstanding feature of NRQCD is that it treats the quarkonium not simply as
a quark-antiquark pair in color-singlet but rather a superposition of Fock
states. Although generally the color-singlet takes a more important role in 
quarkonium production and decays, the other high Fock states may be dominant 
in some cases. Under this framework, one can calculate
the inclusive production and decay rates to any order in strong coupling constants
$\alpha_s$ as well as $v^2$, the relative velocity of heavy quarks inside
the bound state. Just with this mechanism, the authors of Ref.[3-5] have 
successfully explained the $\jpsi(\psi^\prime)$ and $\Upsilon$ production 
surplus problems discovered by {\bf CDF} group.

$\jpsi$ production in $\ee$ annihilation process has been investigated by several authors
within color-singlet model[6-11]. Recently, Braaten and Chen have noted that
a clean signature of color-octet mechanism may be observed in the angular
distribution of $\jpsi$ production near the endpoint region at $\ee$ collider\cite{chen}.
In this paper, we make a thorough discussion about the $\jpsi$ production at $\ee$ colliders,
including color-singlet and color-octet contributions.
The rest of the paper is organized as follows. In Sec.II, we discuss the 
$\jpsi$ production color-singlet processes, including
1) $e^+e^-\to J/\psi gg$; 2) $e^+e^-\to J/\psi c\bar{c}$; 3) $e^+e^-\to q 
\bar q ggJ/\psi$ and $e^+e^-\to q \bar q g\chi_c$ followed by $\chi_c \to J/\psi \gamma$.
We carefully study the characters of each channel. In Sec.III, we study the 
color-octet production mechanism, including
1) $e^+e^-\to J/\psi g$; 2) $\ee\to \jpsi q\bar q$. 
Comparisons of the energy scaling properties as well as energy spectrum of these
processes are made. In Sec.IV, we give a short discussion about the determination of 
color-octet matrix elements, and find that they can be extracted from $\ee$ 
collision experiments at different energy regions. 
Finally, we briefly discuss our results and make a conclusion in Sec.V.

\section{Color-singlet contributions}
~

The leading order color-singlet contributions to direct $\jpsi$ production
include the following processes
\begin{eqnarray}
\ee&\to&\gamma^*\to\jpsi c\bar c,\\
\ee&\to&\gamma^*\to\jpsi gg,\\
\nonumber
\ee&\to&\gamma^*\to q\bar q g^*~~{\rm with}~~g^*\to\jpsi gg,\\
\ea
The relevant Feynman digrams are shown in Fig.1. Fig.1(a) is a quark process,
Fig1.(b) is a gluon process, Fig.1(c) and Fig.1(d) are gluon jets processes.
All these three processes have been calculated separately in literatures[8-11].
Here, we make a comparison of the relative weight of them as a function of c.m. energy
$\sqrt{s}$.

For the quark process, the calculation is straightforward, and we get\cite{cho}
\beq
\label{e1}
\frac{d\sigma(\ee\to\jpsi c\bar c)}{\sigma_{\mu\mu}dzdx_1}=
\frac{32\alpha_s^2 e_c^2}{243}\frac{<{\cal O}_1^{\psi}(^3S_1)>}{m^3}
\frac{\sum\limits_{i=1}^{4}F_i r^i}{(1-x_1)^2(2-z)^2(z+x_1-1)^4}
,
\eeq
where
\beq
\label{e2}
z=\frac{2p\cdot k}{s},~~~x_i=\frac{2p_i\cdot k}{s},~~~r=\frac{m^2}{s},
\eeq
$$\sigma_{\mu\mu}=\sigma_{QED}(\ee\to\mu^+\mu^-).$$
$k$, $p$ and $p_i$ are the momenta of virtual photon $\gamma^*$, $\jpsi$, 
and outgoing parton ($c$ quark in Fig.1(a), gluon in Fig.1(b)), 
respectively; $m$ is the mass of $\jpsi$, which equals to $2m_c$ at 
nonrelativistic approximation;
$<{\cal O}_1^{\psi}(^3S_1)>$ is the color-singlet nonperturbative matrix element;
The functions $F_i$ are defined as
\begin{eqnarray}
\nonumber
F_1&=&2 (x_1 + z - 1)^2 (x_1 - 1)[4 x_1^3 - 6 x_1^2 (z^2 - 2 z + 2) \\
\nonumber
&~&- 2 x_1 (4 z^3 -15  z^2 + 12 z -2) - 11 z^4 + 23 z^3 - 24 z^2 + 8 z + 4],\\
\nonumber
F_2&=&(x_1 + z - 1)[16 x_1^5 + 4 x_1^4 (5 z - 4) + 4 x_1^3(3 z^2 - 4 z - 24)
        + x_1^2(5 z^3 + 16 z^2 - 180 z + 256)\\ 
\nonumber
&~&+x_1 (-7 z^4 + 30 z^3 - 132 z^2 + 360 z - 256) + 11 z^4 - 55 z^3 + 136 z^2 - 200 z + 96],\\
\nonumber
F_3&=&8 (4 x_1^5 - 14 x_1^4 + 31 x_1^3 - 50 x_1^2 + 43 x_1 - 14)
        +4 z (14 x_1^4 - 62 x_1^3 + 149 x_1^2 - 184 x_1 + 83)\\
\nonumber
&~&+2 z^2 (21 x_1^3 - 130 x_1^2 + 269 x_1 - 176)+z^3 (29 x_1^2 - 154 x_1 + 177)+ 4 z^4 (4 x_1 - 11)+5 z^5, \\
F_4&=&3 (4 x_1^3 + 4 x_1^2 z - 4 x_1^2 - 4 x_1 - z^3 + 4 z^2 - 8 z + 4) (2
        x_1 + z - 2).
\end{eqnarray}                                               

After integrating Eq.(\ref{e1}) over $x_1$, and considering the fragmentation
approximation ({\it i.e.}, $\sqrt{s}\gg m_c$, $r\ll 1$), the differential cross
section reads as
\beq
\frac{d\sigma}{dz}(\ee\to\jpsi c\bar c)=2\sigma(\ee\to c\bar c)\times 
        D_{c\to \jpsi}(z),
\eeq
where $D_{c\to \jpsi}(z)$ denotes the quark fragmentation function. Its explicit 
form may be found in Ref.\cite{frag}.

As for the gluon process Fig.1(b), from Ref.\cite{keung} we readily have
\beq
\frac{d\sigma(\ee\to\jpsi gg)}{\sigma_{\mu\mu}dzdx_1}=
\frac{64e_c^2 \alpha_s^2}{81}\frac{<{\cal O}_1^{\psi}(^3S_1)>}{m^3}
r^2 f(z,x_1;r),
\eeq
where 
\begin{eqnarray}
\label{e3}
\nonumber
f(z,x_1;r)&=&\frac{(2+x_2)x_2}{(2-z)^2(1-x_1-r)^2} +
 \frac{(2+x_1)x_1}{(2-z)^2(1-x_2-r)^2}\\ 
 \nonumber
 &+&
 \frac{(z-r)^2-1}{(1-x_2-r)^2(1-x_1-r)^2}
+ \frac{1}{(2-z)^2}\Big(\frac{6(1+r-z)^2}
 {(1-x_2-r)^2(1-x_1-r)^2}\\ 
& + & \frac{2(1-z)(1-r)}{(1-x_2-r)(1-x_1-r)r}
  +\frac{1}{r}\Big ),
\end{eqnarray}
the variables $z,~x_i,~r$ are defined as Eq.(\ref{e2}) and $x_2=2-z-x_1$.

$\jpsi$ production from gluon jets processes as shown in Fig.1(c) and Fig.1(d)
have been calculated in Ref.\cite{hagiwara}, giving 
\beq
\sigma(\ee\to q\bar q g^*;g^*\to\jpsi X)=
        \int\limits_{m^2}^{s}d\mu^2 \sigma(\ee\to q\bar q g^*(\mu))P(g^*\to\jpsi X),
\eeq
where $\mu=m(g^*) $ is the virtuality of the gluon.
$P(g^*\to \jpsi X)$ is the decay distribution function of virtual gluon to $\jpsi$,
which includes the contributions directly from $g^*\to \jpsi gg$
and from E1 transitions of $\chi_c$ ($g^*\to g\chi_c$ followed by
$\chi_c\to \jpsi \gamma$). We can express it as 
\beq
P(g^*\to \jpsi X)=P_S(g^*\to \jpsi gg)+Br(\chi_c\to \jpsi \gamma)P(g^*\to \chi_c g),
\eeq
where
\beq
P_S(g^*\to \jpsi gg)=\frac{10\alpha_s^3}{243\pi}\frac{<{\cal O}_1^{\psi}(^3S_1)>}{m^3}
        \frac{r^2}{\mu^2}\int\limits_{2\sqrt{r}}^{1+r}dz\int\limits_{x_-}^{x_+}
        dx_1 f(z,x_1;r).
\eeq
Here the function $f(z,x_1;r)$ is defined as Eq.(\ref{e3}) with $r=m^2/\mu^2$. 
The integration limits of $x_1$ are
\beq
x_\pm=\frac{1}{2}(2-z\pm\sqrt{z^2-4 r}).
\eeq
Because the $g^*\to \chi_{cJ} g$ processes have the infrared divergence involved,
which are associated with the soft gluon in the final states, we induce
an infrared cutoff to avoid the singularities in $P(g^*\to \chi_{cJ}g$). 
Strictly speaking, the divergences can be cancelled in the framework
of NRQCD (see Ref.\cite{jpma}). Here, we follow the way of Ref.\cite{pw} by
imposing a lower cutoff $\Lambda$ on the energy of the outgoing gluon in the
quarkonium rest frame. Then, the decay distribution functions for $\chi_{cJ}$
can be written as
\ba
P(g^*\to \chi_{c0} g)&=&\frac{8\alpha_s^2}{9\pi}\frac{|R_P^\prime(0)|^2}{\mu^2m^5}
        \frac{r(1-3r)^2}{1-r}\theta(\frac{1}{r}-\frac{\Lambda}{m_c}-1),\\
P(g^*\to \chi_{c1} g)&=&\frac{8\alpha_s^2}{9\pi}\frac{|R_P^\prime(0)|^2}{\mu^2m^5}
        \frac{6r(1+r)}{1-r}\theta(\frac{1}{r}-\frac{\Lambda}{m_c}-1),\\
P(g^*\to \chi_{c2} g)&=&\frac{8\alpha_s^2}{9\pi}\frac{|R_P^\prime(0)|^2}{\mu^2m^5}
        \frac{2r(1=+r+6r^2)}{1-r}\theta(\frac{1}{r}-\frac{\Lambda}{m_c}-1).
\ea
As discussed in Ref.\cite{pw}, the cutoff $\Lambda$ can be set to $m_c$ in order
to avoid the large logarithms in the divergent terms.

With the above formulas, we can evaluate the prompt $\jpsi$ production rates
of color-singlet processes in $\ee$ annihilation at any
energy regions. The results are displayed in Fig.2 and Fig.3, the input
parameters used in the numerical calculations are \cite{braaten}\cite{quigg}
\beq 
 m_u=m_d=m_s=0,~~~m_c=1.5 GeV,~~~m_b=4.9GeV,~~~\alpha_s(2m_c)=0.26,
\eeq
\beq 
 <{\cal O}_1^{\psi}(^3S_1)>=0.73 GeV^3,~~~~~~|R_P^\prime(0)|^2=0.125 GeV^5.
\eeq
The branching ratios of $\chi_{cJ}\to \jpsi \gamma$ taking in the calculations
are \cite{data}
\ba
\nonumber
 Br(\chi_{c0}\to \jpsi \gamma)&=&6.6\times 10^{-3},\\
\nonumber
 Br(\chi_{c1}\to \jpsi \gamma)&=&27.3\%,\\
 Br(\chi_{c2}\to \jpsi \gamma)&=&13.5\%.
\ea

The angular distribution and energy distribution of color-singlet $\jpsi$
production at {\bf CLEO} have been discussed in Ref.\cite{cho}.
Here, we make a comparison of the relative importance of these three color-singlet
processes as a function of c.m. energy $\sqrt{s}$. The result is displayed in Fig.2.
The dotted line demonstrates the contribution from quark process (1), the dash line
is from gluon process (2), the dotted-dash line is from gluon jets processes (3),
and the solid line is the total color-singlet cross section. From this figure, we 
can see that at low energies ($\le 25~GeV$) the gluon process dominates the other 
two processes, at somewhat high energies the quark
process will dominate, and at high enough energies, the gluon jets processes are dominant. 
To see the sensitivity of quark fragmentation approximation to interaction 
energy $\sqrt{s}$, we also plot the line corresponding to the quark 
fragmentation approximation Eq.(7) in the same figure. Obviously, the quark 
process at high energies can be represented by the quark fragmentation 
approximation. The diagrams show that at $\sqrt{s}\ge 70~GeV$ the difference 
between the complete calculation and the fragmentation approximation is less 
than $5\%$. Another striking result is that
the process (3) is negligible at low energies, but at high enough energies 
($\sqrt{s}\ge 200 GeV$) its contribution dominates the other two processes 
and grows with the energy increases.

In Fig.3, We display the energy distributions of quark process, gluon process, 
and their sum at $\sqrt{s}$ to be $10.6~GeV$, $25 GeV$, 
$50 GeV$ and $100~GeV$ respectively. At low energies, such as at {\bf CLEO}
($\sqrt{s}=10.6 GeV$), the energy spectrum of these two processes are both
flat (see Fig.3(a)), however, at high energies the patterns of the energy
spectrum of these two processes are distinct. The quark process is hard and 
the gluon process is soft (see Fig.3(c)-(d)). 
After the relative importance of these two processes is changed with c.m. 
energy, the energy distribution feature of the total cross section of these 
two processes is also changed with c.m. energy. Therefore, at high enough 
energies, the distribution is mainly from quark process, and the total 
spectrum appears hard (see Fig.3(a)-(d)).

\section{Color-octet contributions}
~

The leading order color-octet contributions to direct $\jpsi$ production in 
$\ee$ collision include the following two processes 
\ba
{\rm i)}~\ee&\to&\gamma^*\to q\bar q+c\bar c [\b{8},^3S_1],\\
{\rm ii)}~\ee&\to&\gamma^*\to g+c\bar c [\b{8},{\rm ^{2S+1}L_J}],
\ea
as shown in Fig.4. Here ${\rm ^{2S+1}L_J}$ denotes the states $^1S_0$ and 
$^3P_J$, $q$ represents the $u,~d,~s,~c$ and $b$ quarks.

As the first process shown in Fig.4(a), using the factorization formalism
described in Ref.\cite{bbl}, we can write the differential cross section as
\beq
d\sigma(\ee\to q\bar q\jpsi)=d\hat\sigma(\ee\to q\bar q c\bar c[\b{8},^3S_1])
        <{\cal O}_8^{\psi}(^3S_1)>,
\eeq
where $d\hat\sigma$ represents the short distance coefficient of the process, 
which can be calculated perturbatively. $<{\cal O}_8^{\psi}(^3S_1)>$ corresponds
to the long distance nonperturbative matrix element. It can be treated as free
parameter or evaluated by fitting the theory to experimental data. The result is 
\beq
\frac{d\sigma(\ee\to q\bar q\jpsi)}{\sigma_{\mu\mu}ds_1ds_2}=
        \frac{e_q^2 \alpha_s^2}{12}\frac{<{\cal O}_8^{\psi}(^3S_1)>}{m_c^3}
        \frac{\sum\limits_{i=0}^{3}G_i s_2^i}
        {s^2(s_1-m_q^2)(s-s_1-s_2-m^2-m_q^2)},
\eeq
where
$$s_1=(k-p_1)^2,~~~~s_2=(k-p)^2.$$
In the above, $k$ is the momentum of virtual photon $\gamma^*$, $p$ and $p_1$ are the momenta
of outgoing $\jpsi$ and quark $q$, $m_q$ is the mass of the quark and $e_q$ is 
its charge. The integration limits of $s_1$ and $s_2$ are 
\beq
s_1^\pm=m^2+m_q^2-\frac{1}{2s_2}[s_2(s_2-s+m^2)\pm\lambda^{\frac{1}{2}}
        (s_2,s,m^2)\lambda^{\frac{1}{2}}(s_2,m_q^2,m_q^2)],
\eeq
\beq
s_2^-=4m_q^2,~~~~~~~~s_2^+=(\sqrt{s}-m)^2.
\eeq
$\lambda$ is defined as
$$\lambda(x,y,z)=x^2+y^2+z^2-2xy-2yz-2xz.$$
The functions $G_i$ are
\begin{eqnarray}
\nonumber
G_0&=&3 m^6 m_q^2 + 4 m^2 m_q^6 + 7 m^4 m_q^4+ 2 m_q^8 
        +s ( m^6+ 4 m_q^6  + m^6 +11 m^4 m_q^2 + 16 m^2 m_q^4)\\
\nonumber
&~&+ s^2 (2 m^4 + 11 m^2 m_q^2 +7 m_q^4)+ s^3 (m^2+  3 m_q^2)+
        s_1 ( - m^6 - 6 m^4 m_q^2 - 5 m^4 s\\
\nonumber
&~& - 12 m^2 m_q^4 - 16 m^2 m_q^2 s -5 m^2 s^2 - 8 m_q^6 
        - 12 m_q^4 s - 6 m_q^2 s^2 - s^3)+s_1^2 (3 m^4 \\
\nonumber
&~&+ 12 m^2 m_q^2 + 8 m^2 s+ 12 m_q^4 + 12 m_q^2 s + 3 s^2)
        +4 s_1^3 ( - m^2 - 2 m_q^2 - s)+2 s_1^4, \\
\nonumber
G_1&=& - 5 m^4 m_q^2 - 8 m^2 m_q^4 - 2 s (m^4 + 6 m^2 m_q^2 + 4 m_q^4)
        - s^2 (2 m^2 + 5 m_q^2)\\
\nonumber
&~&+s_1 [m^4 + 4 m^2 m_q^2  + 4 m_q^4 + 4 s(m^2 +m_q^2) + s^2]+
        4 s_1^2 ( - m^2 - 2 m_q^2 - s)+4 s_1^3, \\
\nonumber
G_2&=&3 m^2 m_q^2 + m^2 s - m^2 s_1 + 3 m_q^4 + 3 m_q^2 s - 2 m_q^2 s_1 - 
        s s_1 + 3 s_1^2,\\
G_3&=&s_1- m_q^2 .
\end{eqnarray}

A check can be performed by considering the high energy limit in this process.
At high enough energies, the outgoing quark mass $m_q$ can be neglected. Setting
$m_q=0$, and integrating over $s_1$ and $s_2$, we will obtain
\begin{eqnarray}
\label{a10}
\nonumber
\frac{\sigma(\ee\to\jpsi q\bar q)}{\sigma_{\mu\mu}} & = &
\frac{e_q^2\alpha_s^2(2m_c)}{96}\frac{<O^{\psi}_8(^3S_1)>}
{m^3}\big\{ 5(1-r^2)
-2r \ln r\\ 
\nonumber
&+& \big[2Li_2(\frac{r }{1+r })
 - 2 Li_2(\frac{1}{1+r })\\
&-& 2\ln(1+r )\ln r  + 3\ln r  + \ln^2 r \big]
(1+r)^2 \big\},
\end{eqnarray}
where $Li_2(x)=-\int\limits_0^x dt ~{\rm ln}(1-t)/t$ is the Spence function.
The result here is completely consistent with that in Ref.\cite{z0}, in which 
the charmonium production in $Z^0$ decays through the similar process is 
calculated.

The second process as shown in Fig.4(b) has been calculated in Ref.\cite{chen}, 
giving
\ba
\label{combination}
\sigma(\ee\to\jpsi g)&=&C_s <{\cal O}_8^{\psi}(^1S_0)>+C_p <{\cal O}_8^{\psi}(^3P_0)>,\\
C_s&=&\frac{64\pi^2e_c^2\alpha^2\alpha_s}{3}
        \frac{1-r}{s^2m},\\
C_p&=&\frac{256\pi^2\alpha^2\alpha_s}{9s^2m^3}
        \big [\frac{(1-3r)^2}{1-r}+\frac{6(1+r)}{1-r}+
        \frac{2(1+3r+6r^2)}{1-r}\big ],
\ea
where we have used the approximate heavy quark spin symmetry relations:
\beq
<{\cal O}_8^{\psi}(^3P_J)>\approx (2J+1)<{\cal O}_8^{\psi}(^3P_0)>.
\eeq

Up to now, the color-octet matrix elements $<{\cal O}_8^{\psi}(^3S_1)>$,
$<{\cal O}_8^{\psi}(^1S_0)>$ and $<{\cal O}_8^{\psi}(^3P_0)>$ are determined
only by fitting to the experimental data. 
$<{\cal O}_8^{\psi}(^3S_1)>$ is extracted from hadroproduction process at the 
{\bf Tevatron}[2][3]\cite{braaten}. The results are consistent with the theoretical
anticipation of NRQCD. As for $<{\cal O}_8^{\psi}(^1S_0)>$ and $<{\cal O}_8^{\psi}(^3P_0)>$, 
they have been obtained from both hadroproduction and photoproduction, 
but the results from different processes are incompatible (see Ref.\cite{photon}),
which we will give a detailed discussion in Sec.IV. Here, we tentatively choose 
the color-octet matrix elements obtained in Ref.\cite{cho1}\cite{braaten},
which are consistent with the velocity scaling rules,
\ba
<{\cal O}_8^{\psi}(^3S_1)>&=&1.5\times 10^{-2} GeV^3,\\
<{\cal O}_8^{\psi}(^1S_0)>&=&10^{-2} GeV^3,\\
\frac{<{\cal O}_8^{\psi}(^3P_0)>}{m_c^2}&=&10^{-2} GeV^3.
\ea

Using the input parameters as in Sec.II, we can calculate the intermediate 
color-octet contributions to $\jpsi$ production in $\ee$ collisions. 
>From Eq.(23)-(26) and Eq.(28)-(34), the contributions of these two processes 
to $\jpsi$ production rates as a function of c.m. energy $\sqrt{s}$ may be 
obtained as shown in Fig.5.
The dotted line is from process (ii), the dotted-dash line is from process (i),
the solid line is the sum of them, and the dash line is the total
cross section of color-singlet processes contributions. From this diagram, we can see that
at low energies the dominant process is channel (ii), at higher energies
($\sqrt{s} > 20GeV$) the channel (i) dominates.
It is interesting to note that the color-octet contributions dominate the color-singlet 
contributions at any energy values of $\sqrt{s}$. So if the color-octet production
mechanism is correct and the color-octet matrix elements are not far small
with values used here, $\jpsi$ production in $\ee$ collision mainly comes from 
color-octet contributions.

Another important character of the color-octet $\jpsi$ production in $\ee$ collision is
its energy spectrum. The energy spectrum of process (ii) has been discussed in
Ref.\cite{chen}. In this process, $\psi$ 's energy mainly fixes at the endpoint
region which equals half of the c.m energy, its width is about $(\frac{2m_c}
{\sqrt{s}}\times 500)~MeV$. In contrast, the energy spectrum of process (i)
is very soft. As shown in Fig.6, the energy distribution of $\jpsi$ production
in process (i) is mainly from low energy region even at high c.m. energies (
$\ge 500 GeV$). Numerical result shows, the mean energy value of $\jpsi$ from
process (i) is about $10GeV\sim 20GeV$ at any energy value of $\sqrt{s}$
less than $1000 GeV$.

\section{Color-octet matrix elements} 
~

Unlike color-singlet matrix elements which associate with the quarkonium radial
wave function at the origin in the nonrelativistic limit, and can be 
calculated by potential model\cite{quigg}, the color-octet matrix elements
are unknown. They can be extracted from experimental data or 
from lattice QCD calculations. Before lattice QCD giving out the results, 
the color-octet matrix elements are determined only 
by fitting the theoretical prediction to experimental data.
As done in the literatures[2-4]\cite{braaten}\cite{photon}\cite{fixed},
the color-octet matrix elements are extracted from experimental data of
$\jpsi$ production in hadron collisions and $e^-p$ collisions. In these
processes, the production mechanism is associated with the structure function
of hadrons, therefore there are still left a lot of uncertainties, e.g.
the higher twist effects, which we can not handle clearly now. Therefore, 
the extraction of octet matrix elements from these different processes 
certainly exists large theoretical uncertainties, 
even their results are not consistent with each other\cite{photon}\cite{fixed}.
In contrast, the mechanism of $\jpsi$ production in $\ee$ annihilation process
is much clearer than those hadron processes discussed above. The parton structure
is simpler, and there is no higher twist effects to be considered, so the 
theoretical uncertainty is much smaller. Following, we discuss the possibility of 
extracting elements $<{\cal O}_8^{\psi}(^3S_1)>$, $<{\cal O}_8^{\psi}(^1S_0)>$ and 
$<{\cal O}_8^{\psi}(^1S_0)>$ from $\ee$ annihilation experiments. 

>From the results of Sec.III, the color-octet contributions dominate the 
color-singlet contributions at any energy regions. The total cross section
of $\jpsi$ production in $\ee$ collision is sensitive to the color-octet
matrix elements.
So, we can precisely extract them from fitting the theoretical prediction 
to the experimental data. Furthermore, there is an another important 
feature of color-octet $\jpsi$ production in $\ee$ process, that
the relative importance of these octet processes are distinct at different
energy regions. At high energies, the element $<{\cal O}_8^{\psi}(^3S_1)>$      
is important, while at low energies, $<{\cal O}_8^{\psi}(^1S_0)>$ and $<{\cal O}_8^{\psi}(^3P_0)>$
are important. So, we can extract them separately from different energy
experiments.

$<{\cal O}_8^{\psi}(^1S_0)>$ was determined previously from hadroproduction
at the {\bf Fermilab Tevatron}. In Ref.\cite{braaten}, the authors use the
gluon fragmentation approximation at high $P_\perp$ to fit the experimental
data. They obtain
\beq
\label{frag}
<{\cal O}_8^{\psi}(^3S_1)>=1.5\times 10^{-2} GeV^3.
\eeq
In Ref.\cite{cho1}, the authors extend out the gluon fragmentation region, 
and take a global fitting to the experimental data (at
$P_\perp > 5GeV$) including $<{\cal O}_8^{\psi}(^3S_1)>$, 
$<{\cal O}_8^{\psi}(^1S_0)>$ and $<{\cal O}_8^{\psi}(^3P_0)>$ contributions,
get
\beq
\label{cho2}
<{\cal O}_8^{\psi}(^3S_1)>=6.6\times 10^{-3} GeV^3.
\eeq
At {\bf LEP II} energy region ($\sqrt{s}\sim 160 GeV$), the color-octet 
production cross section $\sigma_8$ is about 6 times larger than the 
color-singlet process cross section $\sigma_1$, and the former mainly comes from the 
contribution of $<{\cal O}_8^{\psi}(^3S_1)>$ (the contributions from 
$<{\cal O}_8^{\psi}(^1S_0)>$ and $<{\cal O}_8^{\psi}(^3P_0)>$ are small enough
and may be neglected, see Fig.5). So, $<{\cal O}_8^{\psi}(^3S_1)>$ can be 
extracted by precisely measuring the total cross section of $\jpsi$ production in $\ee$
annihilation at this energy region.

As for the elements $<{\cal O}_8^{\psi}(^1S_0)>$ and $<{\cal O}_8^{\psi}(^3P_0)>$, 
the situation is more complicated. In previous studies, the values of them were
extracted from the experimental data of hadroproduction at high $P_\perp$,
photoproduction at forward direction and fixed-target hadroproduction.
In Ref.\cite{cho1}, a global fitting to all $P_\perp$ region data shows that
at low $P_\perp$ boundary the theoretical prediction
is dominated by the contributions from $<{\cal O}_8^{\psi}(^1S_0)>$ and $<{\cal O}_8^{\psi}(^3P_0)>$,
and the fitted result is
\beq
\label{teva}
<{\cal O}_8^{\psi}(^1S_0)> +\frac{3}{m_c^2}<{\cal O}_8^{\psi}(^3P_0)>
        =6.6\times 10^{-2}.
\eeq
However, the studies of photoproduction at $e^-p$ collisions show that the 
matrix element values in above equation may be overestimated\cite{photon}, 
and the authors obtain an another linear combination of these two elements 
\beq
\label{photo}
<{\cal O}_8^{\psi}(^1S_0)> +\frac{7}{m_c^2}<{\cal O}_8^{\psi}(^3P_0)>
        =2.0\times 10^{-2}.
\eeq
Eqs.(\ref{teva}) and (\ref{photo}) are incompatible. Furthermore, the fixed-target 
result\cite{fixed} gives the same argument for the matrix elements values in Eq.(\ref{teva}),
and gives
\beq
<{\cal O}_8^{\psi}(^1S_0)> +\frac{7}{m_c^2}<{\cal O}_8^{\psi}(^3P_0)>
        =3.0\times 10^{-2}.
\eeq
This problem may be further clarified in $\jpsi$ production in $\ee$ collision 
experiment. Because at low energies the $\jpsi$ production 
dominantly comes from the color-octet $^1S_0$ and $^3P_J$ subprocesses (see Fig.5),
and where the associated color-octet matrix elements are $<{\cal O}_8^{\psi}(^1S_0)>$ and 
$<{\cal O}_8^{\psi}(^3P_0)>$. We can extract an another linear combination of
these two elements by fitting to the experimental data at this energy regions.
Furthermore, the coefficients in front of these two elements in the
combination are different at different c.m. energy, because the relative
importance of these two subprocesses is changed with energy.
So we can extract different combinations from different energy experiments. 
Here we choose two typical c.m. energy values, $\sqrt{s}=4.6 GeV$ (at {\bf BEPC}) 
and $10.6 GeV$ (at {\bf CLEO}), to see what combination of the two elements can 
be extracted from the experiment. At $\sqrt{s}=4.6 GeV$, from Eq.(\ref{combination}) we get
\beq
\Delta_8=0.065<{\cal O}_8^{\psi}(^1S_0)>+\frac{<{\cal O}_8^{\psi}(^3P_0)>}{m_c^2},
\eeq
and at $\sqrt{s}=10.6 GeV$, the combination is
\beq
\Delta_8=0.26<{\cal O}_8^{\psi}(^1S_0)>+\frac{<{\cal O}_8^{\psi}(^3P_0)>}{m_c^2}.
\eeq
>From these two equations, the individual value of these two elements may be
extracted.

\section{Conclusions}
~

In this paper, we have calculated the direct $\jpsi$ production in $\ee$ annihilation
including color-singlet and color-octet contributions. We have studied the energy
scaling properties as well as energy spectrum of all the production 
processes. The numerical result shows
that the color-octet contributions dominate the color-singlet contributions 
at any c.m. energy scales and the energy spectrum of these two color-octet 
processes are distinct. That enables us to further carefully study the 
properties of color-octet $\jpsi$ production in $\ee$ collision experiment 
such as the angular distributions {\it et al.}. 

In this paper, we have concentrated on the $\jpsi$ production only through the 
virtual photon in $\ee$ collisions. The contributions from the $Z^0$ boson 
at high energies should be included and will be considered elsewhere.


Because in $\ee$ processes $\jpsi$ production has much
smaller theoretical uncertainty THAN IN HADRONIC $\jpsi$ PRODUCTION
PROCESSES, 
it can be used to extract the color-octet matrix elements precisely.
At high energies color-octet $^3S_1$ subprocess is dominant and the element
$<{\cal O}_8^{\psi}(^3S_1)>$ can be extracted from experimental data in
this energy region. At low energies color-octet $^1S_0$ and $^3P_J$
subprocesses will dominate, one can extract
two sets of linear combinations of the elements $<{\cal O}_8^{\psi}(^1S_0)>$
and $<{\cal O}_8^{\psi}(^3P_0)>$ from different energy experiments, e.g. at 
{\bf BEPC} and {\bf CLEO} energy regions. 
>From these two combination equations, the individual values of the 
matrix elements can be obtained separately. This really provides a strong
motivation in experiment to extract the color-octet matrix elements in $\ee$
colliders at now reaching energy. 
In conclusion, in $\ee$ annihilation experiment, $\jpsi$ production 
signature provides an another criterion in testing the color-octet signals 
and the NRQCD scaling rules.

\vskip 1cm
\begin{center}
\bf\large\bf{Acknowledgements}
\end{center}

One of us (F.Yuan) thanks the staff of the Physics Department Computer Center (Room 540)
for their kind help.
This work was supported in part by the National Natural Science Foundation
of China, the State Education Commission of China and the State Commission
of Science and Technology of China.


\newpage
\centerline{\bf \large Figure Captions}
\vskip 2cm
\noindent
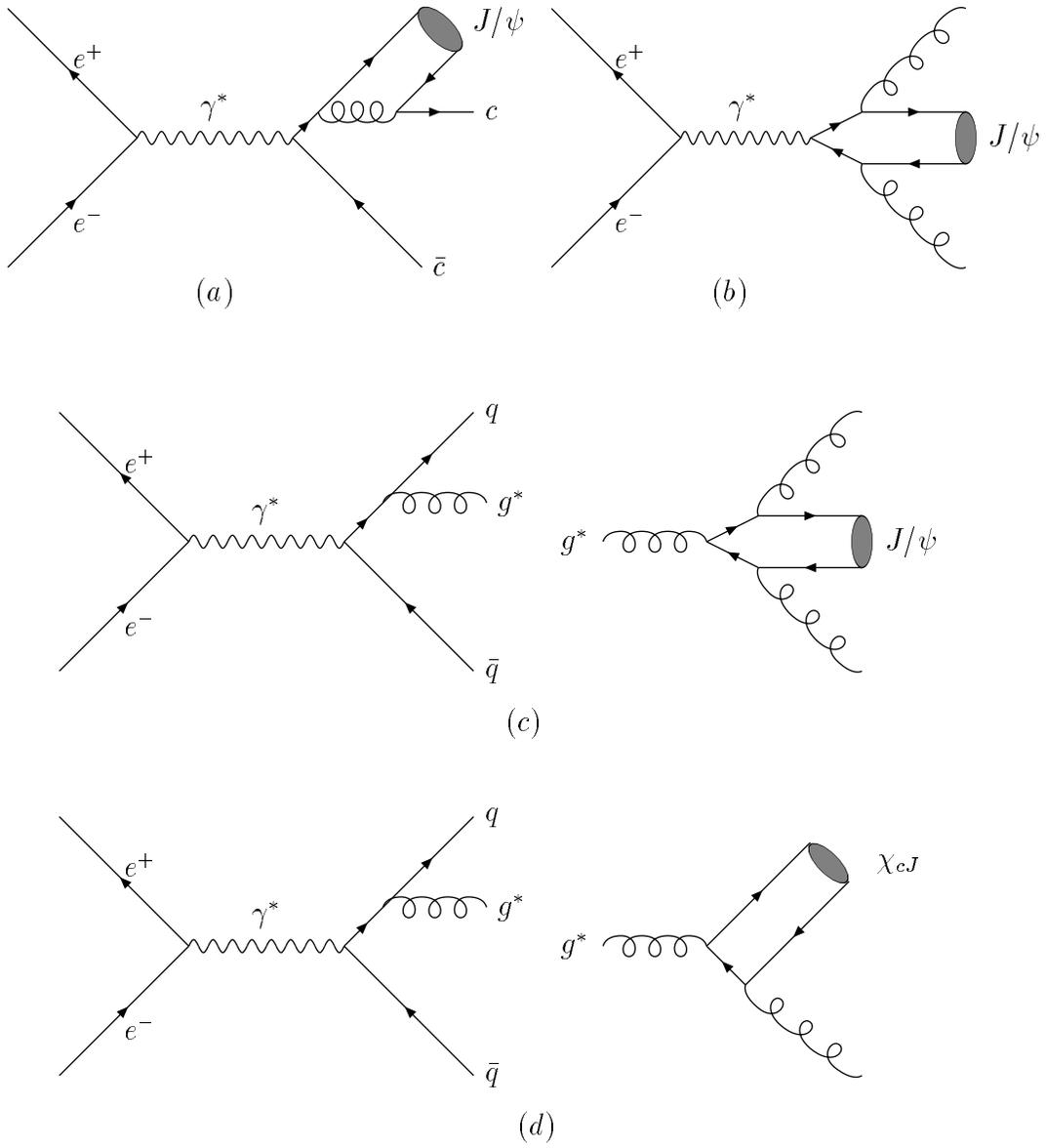
Fig.1. Feynman diagrams for direct $\jpsi$ color-singlet production processes
in $\ee$ annihilation. (a) quark process $\ee\to\jpsi c \bar c$; (b) gluon process 
$\ee\to\jpsi gg$; (c) gluon jets process $\ee\to q\bar q g^*$ with $g^*\to
\jpsi gg$; (d) $\chi_c$ production from gluon jets in $\ee\to q\bar q g^*$ with
$g^*\to \chi_c g$.

\noindent
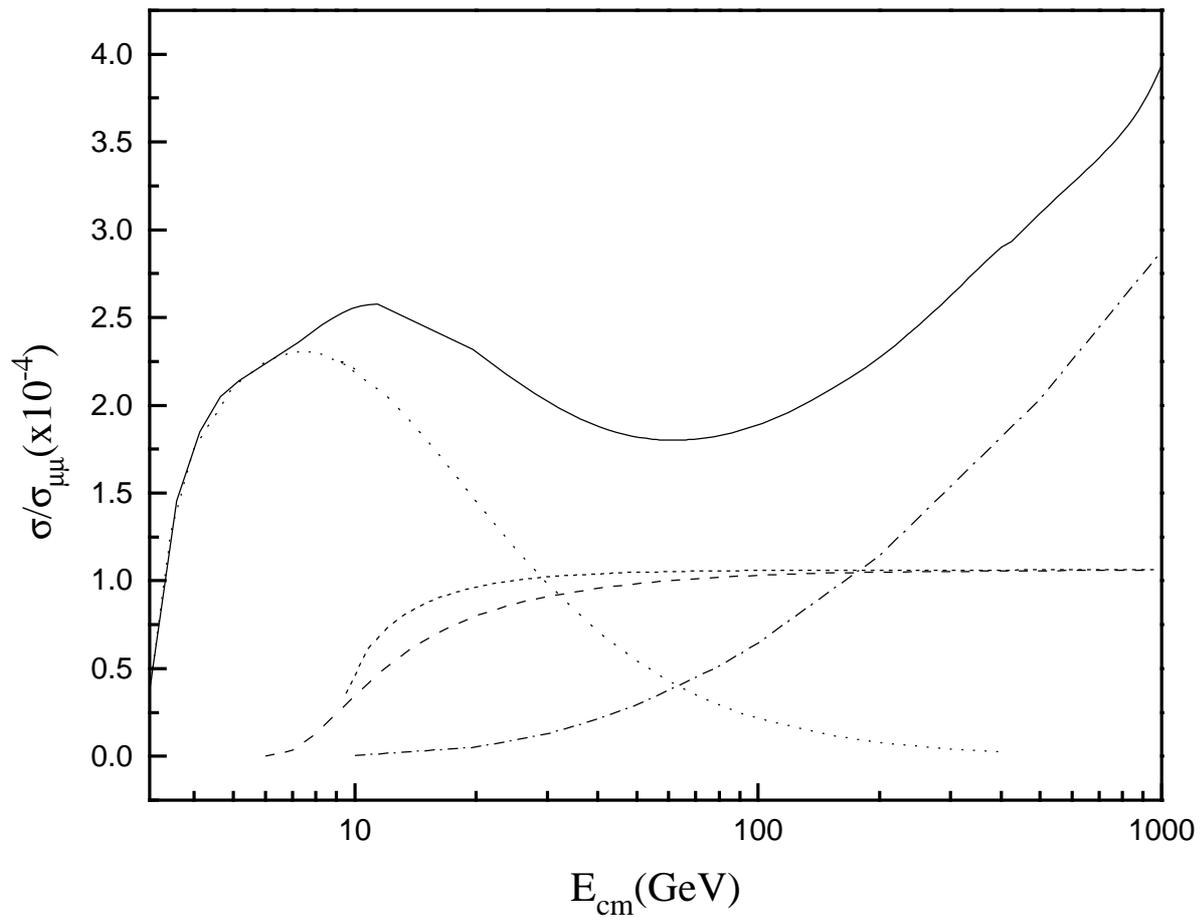
Fig.2. Color-singlet cross section vs c.m. energy. Dotted line illustrates 
the quark process, dash line shows the gluon process, dotted-dash line comes 
from gluon jets, and the sum of the three processes is plotted as solid line. The quark fragmentation
approximation is also shown as short-dashed curve.

\noindent
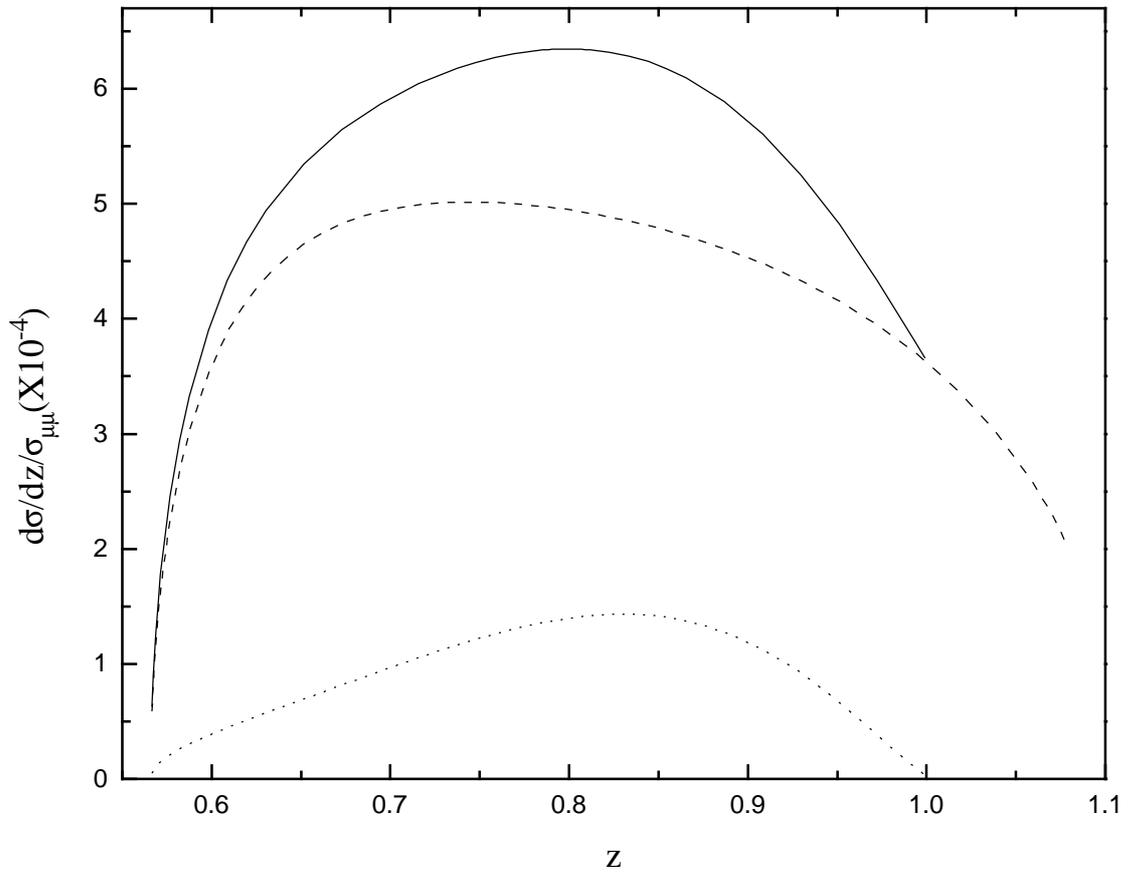
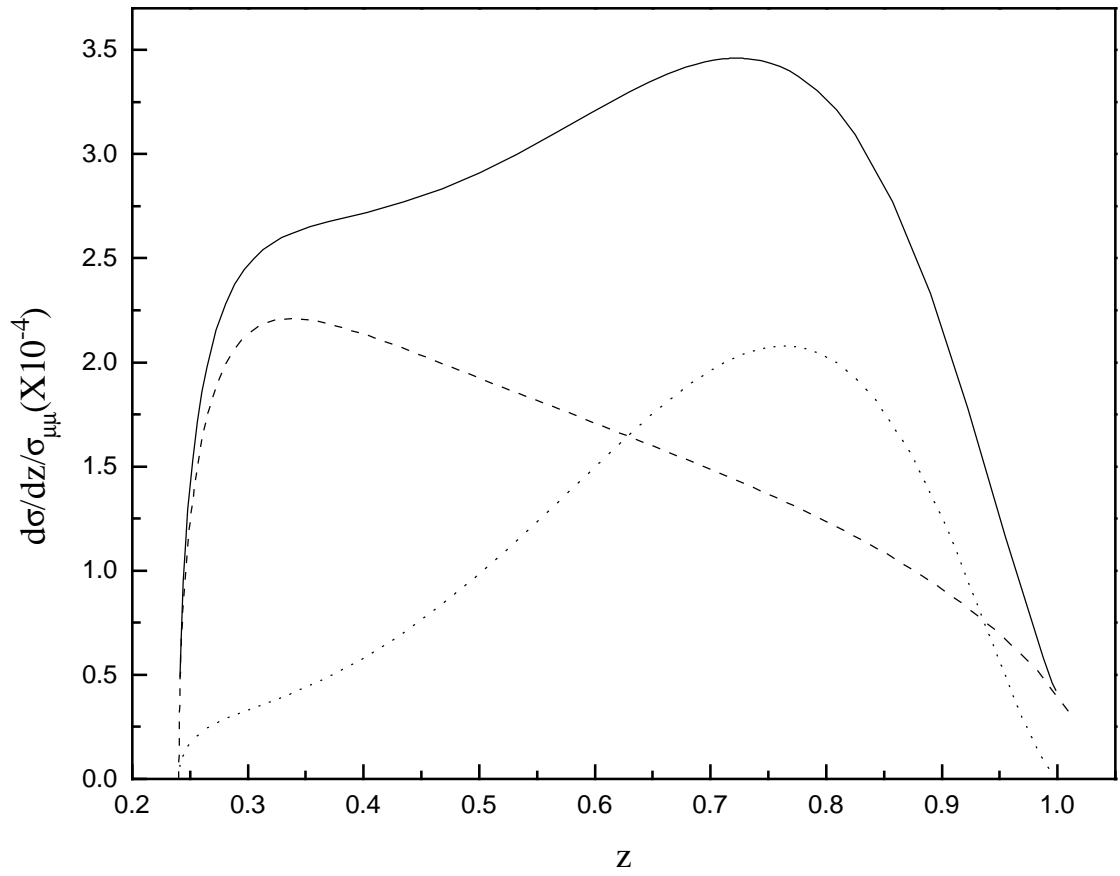
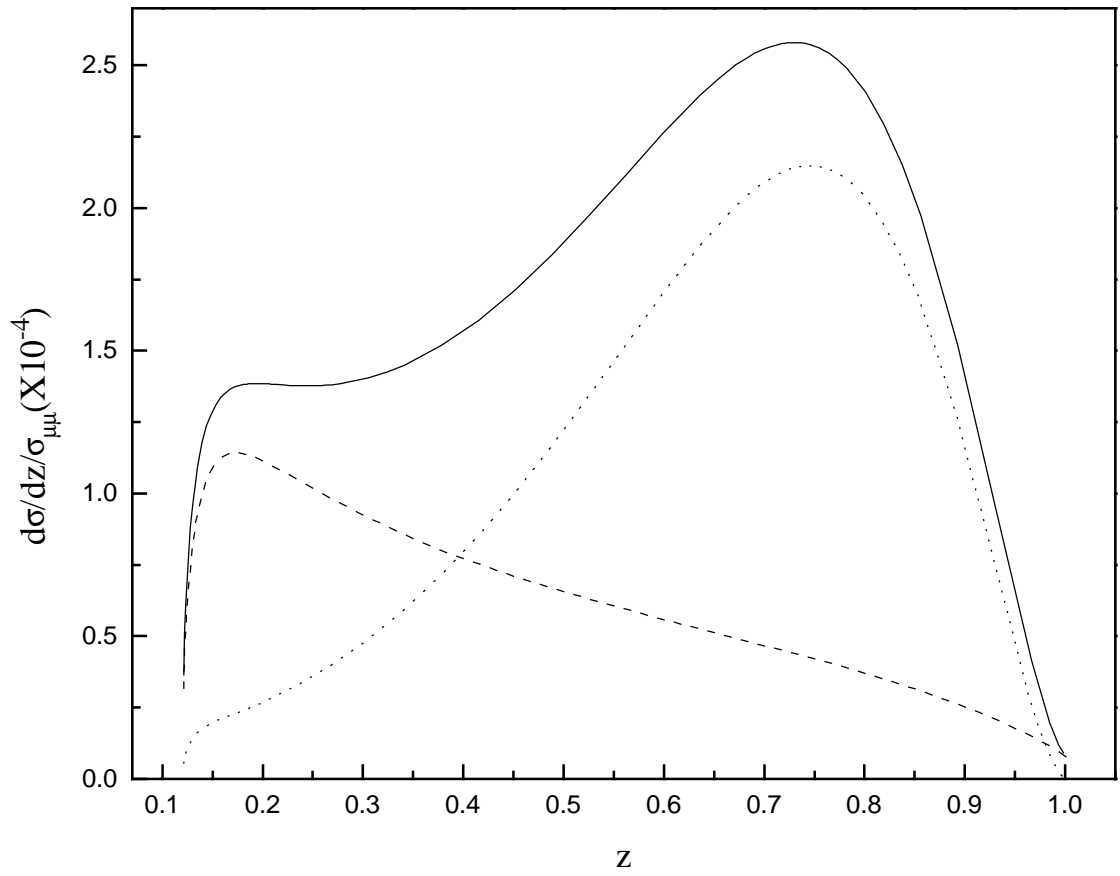
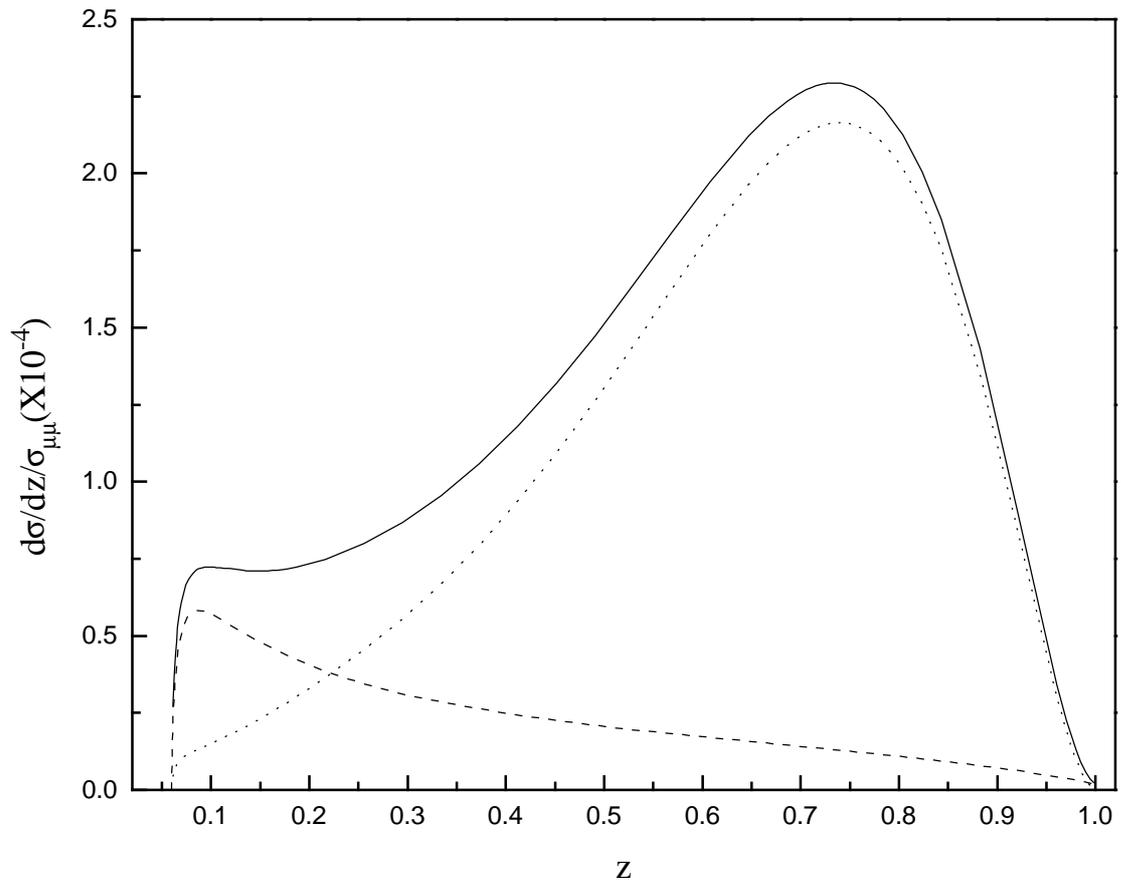
Fig.3. Color-singlet energy distribution $d\sigma/dz$ as function of $z$ at different c.m.
energy. The distributions from quark process (dotted line), gluon process (dashed line)
along with the sum of them (solid line) at (a) $\sqrt{s}=10.6 GeV$;
(b) $\sqrt{s}=25 GeV$; (c) $\sqrt{s}=50 GeV$; (d) $\sqrt{s}=100 GeV$.

\noindent
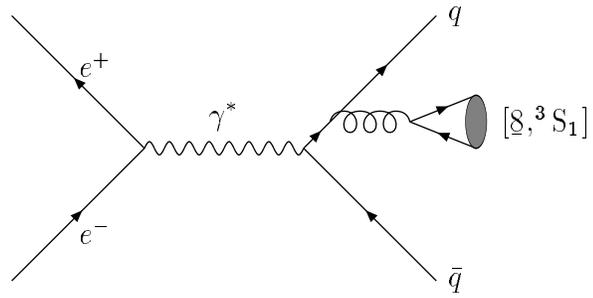
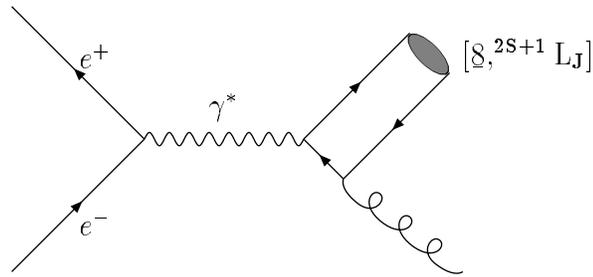
Fig.4. Feynman diagrams for direct $\jpsi$ color-octet production processes
in $\ee$ annihilation. (a) $\ee\to q\bar q \jpsi$; (b) $\ee\to \jpsi g$.

\noindent
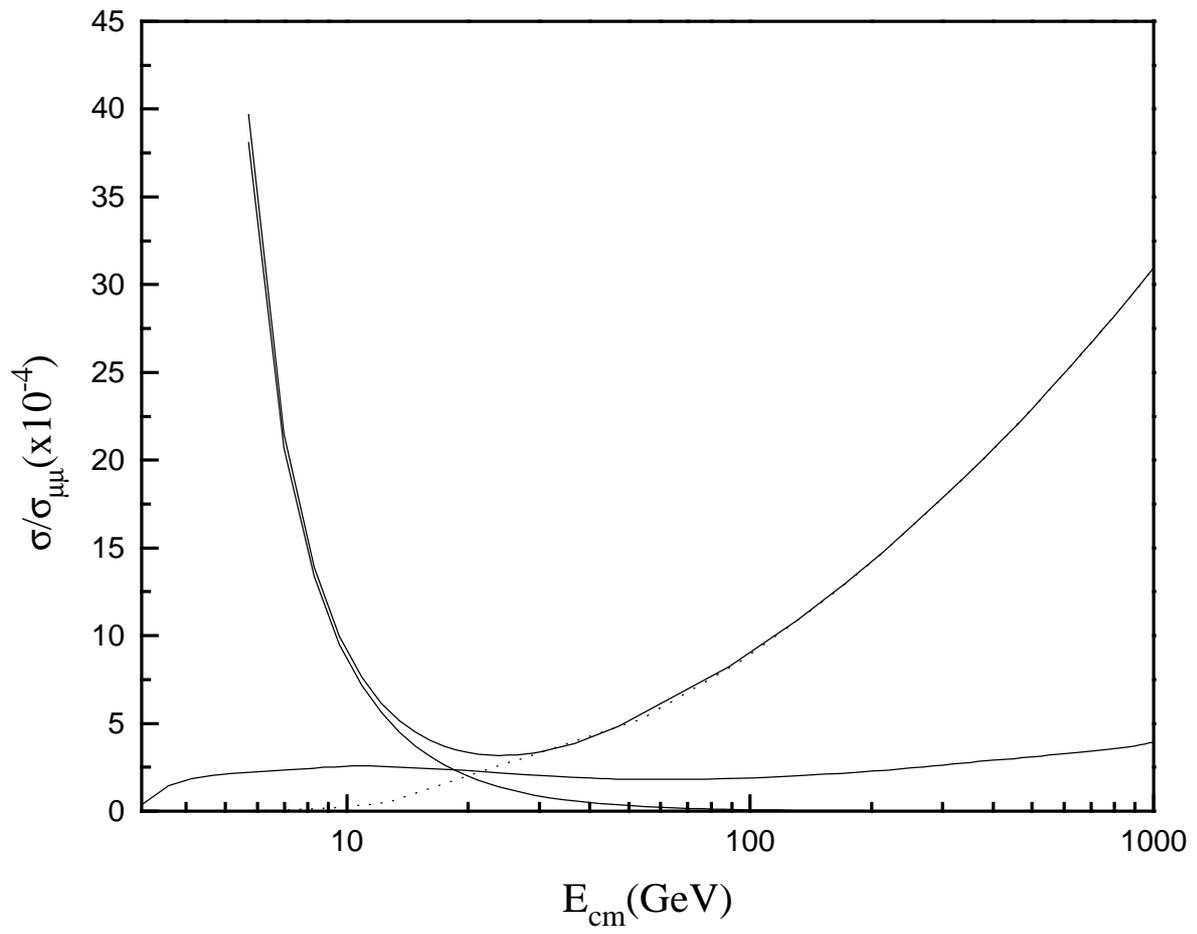
Fig.5. Color-octet cross section vs c.m. energy. Process (i) contribution is 
represented as the dotted-dash line, process (ii) as the dotted line, and 
the sum as the solid line. The color-singlet contribution is drawn as dashed line.

\noindent
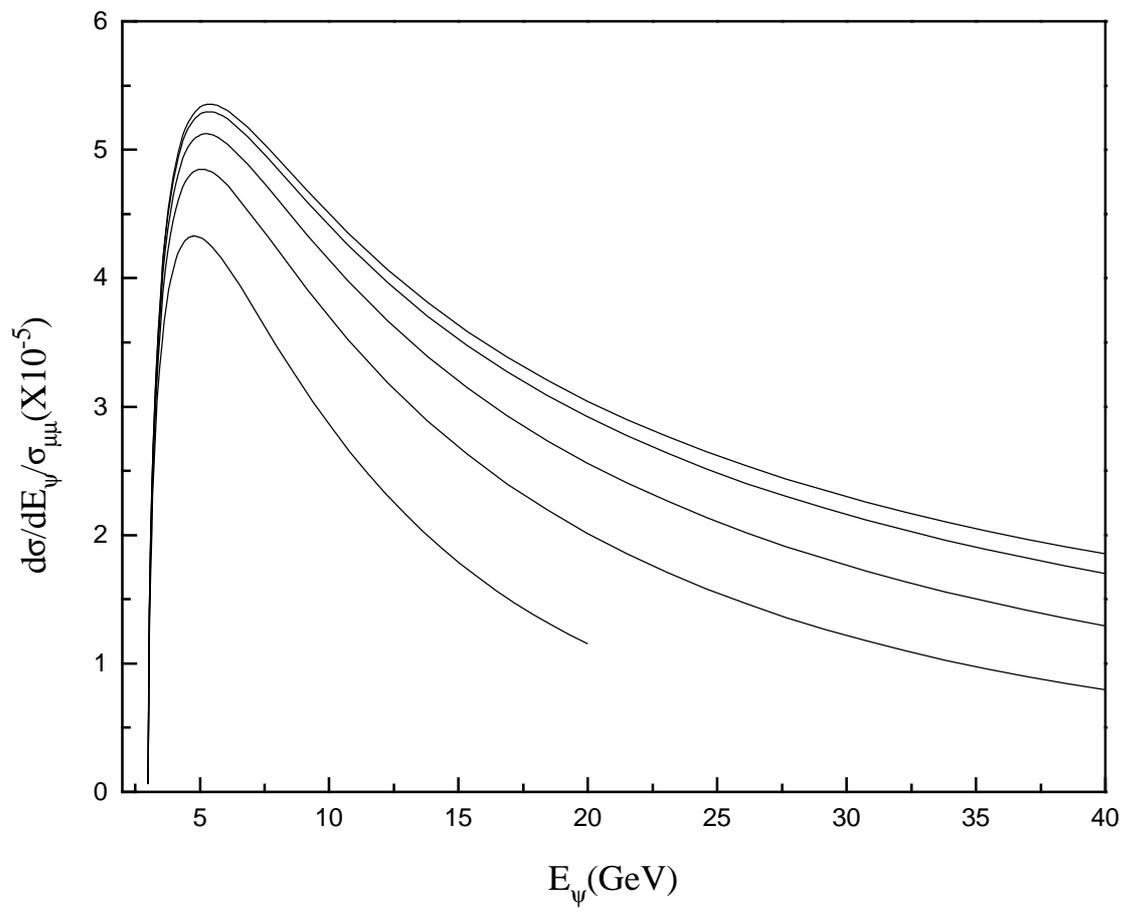
Fig.6. Energy spectrum of produced $\jpsi$ in color-octet  process (i) at different
c.m. energies. From up to down, the curves represent $\sqrt{s}=1000GeV$, $500GeV$, 
$200GeV$, $100GeV$, and $50GeV$ respectively.

\end{document}